\documentclass[aps,twocolumn,showpacs,superscriptaddress,floatfix,longbibliography]{revtex4-1}
\usepackage{amsmath,amssymb}
\usepackage{graphicx}
\usepackage{epsfig}
\usepackage{color}
\usepackage{rotating}
\usepackage{bm}
\usepackage{setspace}

\begin{document}
\title{Observation of Coulomb blockade and Coulomb staircases in superconducting Pr$_{0.8}$Sr$_{0.2}$NiO$_{2}$ films}
\author{Rui-Feng Wang}
\affiliation{State Key Laboratory of Low-Dimensional Quantum Physics, Department of Physics, Tsinghua University, Beijing 100084, China}
\author{Yan-Ling Xiong}
\affiliation{State Key Laboratory of Low-Dimensional Quantum Physics, Department of Physics, Tsinghua University, Beijing 100084, China}
\author{Hang Yan}
\affiliation{State Key Laboratory of Low-Dimensional Quantum Physics, Department of Physics, Tsinghua University, Beijing 100084, China}
\author{Xiaopeng Hu}
\affiliation{State Key Laboratory of Low-Dimensional Quantum Physics, Department of Physics, Tsinghua University, Beijing 100084, China}
\author{Motoki Osada}
\affiliation{Stanford Institute for Materials and Energy Sciences, SLAC National Accelerator Laboratory, Menlo Park, California 94025, United States}
\affiliation{Department of Applied Physics, Stanford University, Stanford, California 94305, United States}
\author{\\Danfeng Li}
\email[]{danfeng.li@cityu.edu.hk}
\affiliation{Department of Physics, City University of Hong Kong, Kowloon, Hong Kong SAR 999077, China}
\author{Harold Y. Hwang}
\affiliation{Stanford Institute for Materials and Energy Sciences, SLAC National Accelerator Laboratory, Menlo Park, California 94025, United States}
\affiliation{Department of Applied Physics, Stanford University, Stanford, California 94305, United States}
\author{Can-Li Song}
\email[]{clsong07@mail.tsinghua.edu.cn}
\affiliation{State Key Laboratory of Low-Dimensional Quantum Physics, Department of Physics, Tsinghua University, Beijing 100084, China}
\affiliation{Frontier Science Center for Quantum Information, Beijing 100084, China}
\author{Xu-Cun Ma}
\email[]{xucunma@mail.tsinghua.edu.cn}
\affiliation{State Key Laboratory of Low-Dimensional Quantum Physics, Department of Physics, Tsinghua University, Beijing 100084, China}
\affiliation{Frontier Science Center for Quantum Information, Beijing 100084, China}
\author{Qi-Kun Xue}
\affiliation{State Key Laboratory of Low-Dimensional Quantum Physics, Department of Physics, Tsinghua University, Beijing 100084, China}
\affiliation{Frontier Science Center for Quantum Information, Beijing 100084, China}
\affiliation{Southern University of Science and Technology, Shenzhen 518055, China}

\begin{abstract}
Motivated by the discovery of superconductivity in the infinite-layer nickelate family, we report an experimental endeavor to clean the surface of nickelate superconductor Pr$_{0.8}$Sr$_{0.2}$NiO$_{2}$ films by Ar$^+$ ion sputtering and subsequent annealing, and we study their electronic structures by cryogenic scanning tunneling microscopy and spectroscopy. The annealed surfaces are characterized by nano-sized clusters and Coulomb staircases with periodicity inversely proportional to the projected area of the nanoclusters, consistent with a double-barrier tunneling junction model. Moreover, the dynamical Coulomb blockade effects are observed and result in well-defined energy gaps around the Fermi level, which correlate closely with the specific configuration of the junctions. These Coulomb blockade-related phenomena provide an alternative plausible cause of the observed gap structure that should be considered in the spectroscopic understanding of nickelate superconductors with the nano-clustered surface.
\end{abstract}
\maketitle

\section{INTRODUCTION}
The recent discovery of the infinite-layer nickelate family R$_{1-x}$(Sr, Ca)$_x$NiO$_2$ (R = Nd, Pr, La) has provided a fascinating platform for exploring electronic correlation and superconductivity in complex oxide materials \cite{li2019superconductivity, osada2020superconducting, zeng2022superconductivity, osada2021nickelate}. Unlike their cuprate counterparts \cite{keimer2015quantum}, the nickelates exhibit a distinct phase diagram without an insulating parent state \cite{li2020superconducting, zeng2020phase, osada2020phase, lee2022character},  and are considered to possess a multiorbital electronic structure \cite{hu2019two, botana2020similarities, choi2020role}. Although surface-sensitive experimental techniques have proved powerful to clarify the electron pairing symmetry and inter-orbital interaction in high-temperature ($T_c$) superconductors \cite{fischer2007scanning, hoffman2011spectroscopic}, they turn out to be challenging for the nickelate films because of the mandatory topotactic reduction process using CaH$_2$, which may significantly degrade the top surface. A recent scanning tunneling microscopy/spectroscopy (STM/STS) study showed the strange coexistence of a V-shaped energy gap and a fully opened energy gap on a nano-clustered Nd$_{1-x}$Sr$_x$NiO$_2$ surface after a long-time vacuum annealing \cite{gu2020single}. Despite several theoretical proposals \cite{gu2021superconductivity, adhikary2020orbital, wang2020distinct, wu2020surface, choubey2021electronic},  the origin of the two different types of energy gaps and the pairing symmetry of nickelate superconductors remain mysterious.

As is well known, the geometry of an object can profoundly affect the electronic properties as its dimensions are reduced to that comparable to characteristic length scales \cite{perenboom1981electronic, likharev1988correlated}. Specifically, if the charging energy $e^2/2C$ ($e$ is the electron charge and $C$ is the capacitance) of a nano-sized object is larger than the energy of thermal fluctuations ($k_BT$) ($k_B$ is the Boltzmann constant), the effect of single-electron tunneling (SET) arises. In widely studied systems such as metal nanoparticles \cite{zhang2005tunneling, kano2015nanoparticle, schonenberger1992single, dubois1996coulomb, qin2020coupling} and discontinuous films \cite{yuan2020observation}, a double-barrier tunneling junction (DBTJ) model based on the orthodox theory provides a fairly good description of the experimental spectra \cite{hanna1991variation, averin1991theory, amman1991analytic}. Controlled by the impedance of the internal junctions, the tunneling current often exhibits equally spaced Coulomb staircases with increasing bias voltage. On the other hand, in a single junction dominating system, quantum fluctuations exert a significant influence on the SET and lead to a dynamic Coulomb blockade (DCB) \cite{delsing1989effect, devoret1990effect}, yielding an energy gap near the Fermi level ($E_F$) \cite{brun2012dynamical, senkpiel2020dynamical}. Here, we report such behaviors on the surface of superconducting Pr$_{0.8}$Sr$_{0.2}$NiO$_2$ (PSNO) films. Our results call for a more comprehensive understanding of the gap-like features of tunneling spectra on the nano-clustered surface of nickelate superconductors.

\section{METHODS}
The infinite-layer PSNO films were prepared on SrTiO$_3$ (STO) substrates by reducing the precursor Pr$_{0.8}$Sr$_{0.2}$NiO$_3$ thin films grown by the pulsed laser deposition, as detailed elsewhere \cite{osada2020superconducting}. Afterward, the samples were  $ex$ $situ$ transferred to our UHV chamber connected to a Unisoku USM 1300 $^3$He STM system. Prior to STM measurements, we cleaned the samples with Ar$^+$ ion sputtering at energies of 500$\sim$1000 eV for 10$\sim$45 minutes under a pressure of $1\times10^{-5}$ Torr and then annealed them in UHV to improve the crystalline quality at varied temperatures. Unless otherwise specified, the STM measurements were conducted at 0.4 K with a polycrystalline PtIr tip, which was cleaned by $e$-beam heating in UHV and calibrated on MBE-grown Ag/Si(111) films. Tunneling spectra were measured using a standard lock-in technique with a small bias modulation at 931 Hz.

\section{RESULTS AND DISCUSSIONS}
\begin{figure}
\includegraphics{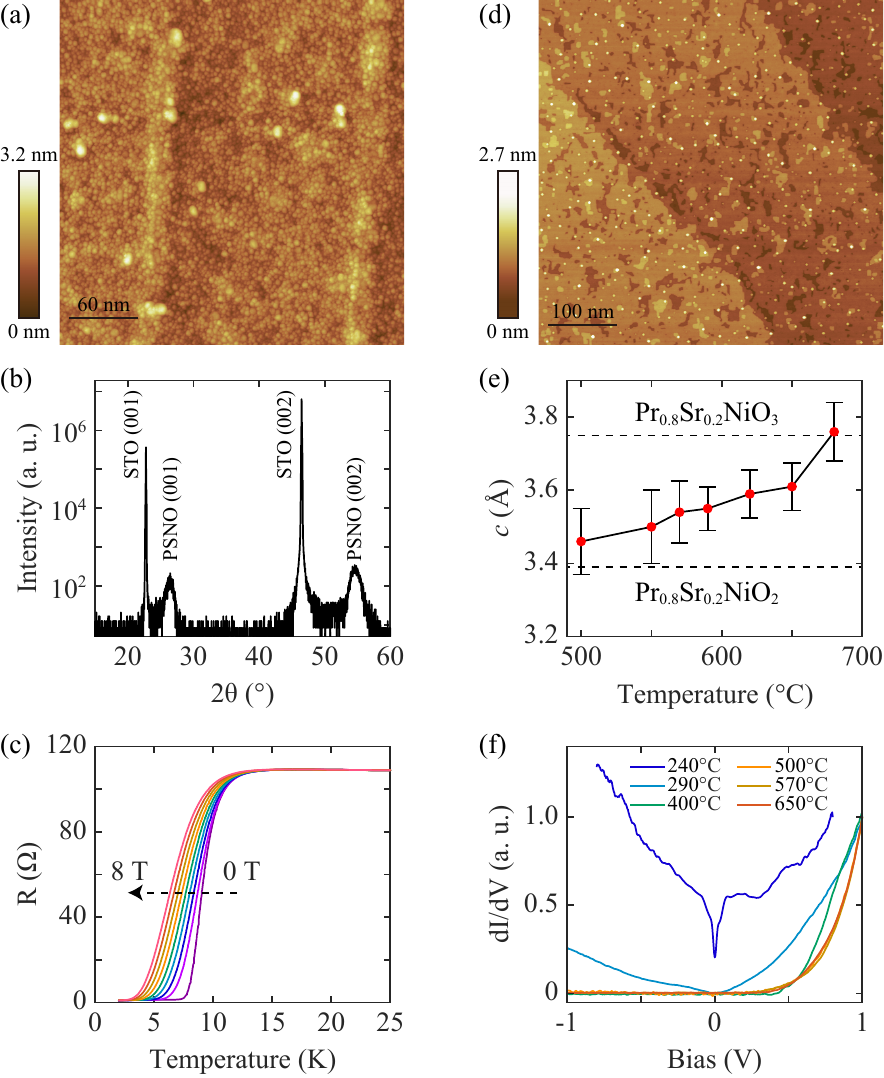}
\caption{\label{fig:1}(a) STM topographic image of PNSO after Ar$^+$ sputtering and annealing at 200$^{\circ}\rm{C}$ (300 nm $\times$ 300 nm, $V$ = -3.0 V, $I$ = 10 pA). (b) XRD pattern (wavelength of the X-ray: 1.5406 Å) of PNSO after treatments at 200$^{\circ}\rm{C}$, showing the infinite-layer PSNO phase. (c) Temperature dependent resistance curves of the PSNO sample in (b) under varied magnetic fields. The superconducting onset temperature of 10.8 K is similar to the films before Ar$^+$  sputtering and annealing. (d) STM topographic image of PNSO after Ar$^+$ sputtering and annealing at 650$^{\circ}\rm{C}$ (500 nm $\times$ 500 nm, $V$ = 2.7 V, $I$ = 10 pA). (e) Step heights of PSNO films as a function of annealing temperatures. Each point comes from the measurements of 20 profiles, and the error bar indicates the standard deviation. (f) Spatially averaged d$I$/d$V$ spectra measured at 4.2 K on the surface of PSNO samples after various annealing temperatures.}
\end{figure}

Figure~\ref{fig:1}(a) shows a typical STM topographic image of the PSNO films upon sputtering at 500 eV for 10 minutes and annealing at 200$^{\circ}\rm{C}$ for 1 hour. The film exhibits a corrugated surface covered with nano-sized clusters. This topography is highly reproducible for all PSNO samples under similar treatments. The samples retain the infinite-layer crystal structure and bulk superconductivity with a $T_c^{onset}$ of 10.8 K (the temperature at which the resistance reduces to 90\% of the value at 20 K), as confirmed by X-ray diffraction (XRD) [Fig.~\ref{fig:1}(b)] and macroscopic transport measurements [Fig.~\ref{fig:1}(c)]. This nano-clustered topography is little affected by further Ar$^+$ ion sputtering.  However, it undergoes an evident transformation into an atomically flat step-terrace structure as the annealing temperature is increased above 500$^{\circ}\rm{C}$, as shown in Fig.~\ref{fig:1}(d). The step height gradually changes from near the $c$-axis length of the PSNO phase (3.39 Å) to that of the Pr$_{0.8}$Sr$_{0.2}$NiO$_3$ phase (3.75 Å) with increasing temperature [Fig.~\ref{fig:1}(e)]. Meanwhile, the tunneling spectra change from metallic characteristics to insulating ones [Fig.~\ref{fig:1}(f)]. We therefore focus on the spectral measurements on the nano-clustered surface after low-temperature annealing (180$\sim$240$^{\circ}\rm{C}$).

\begin{figure}
\includegraphics{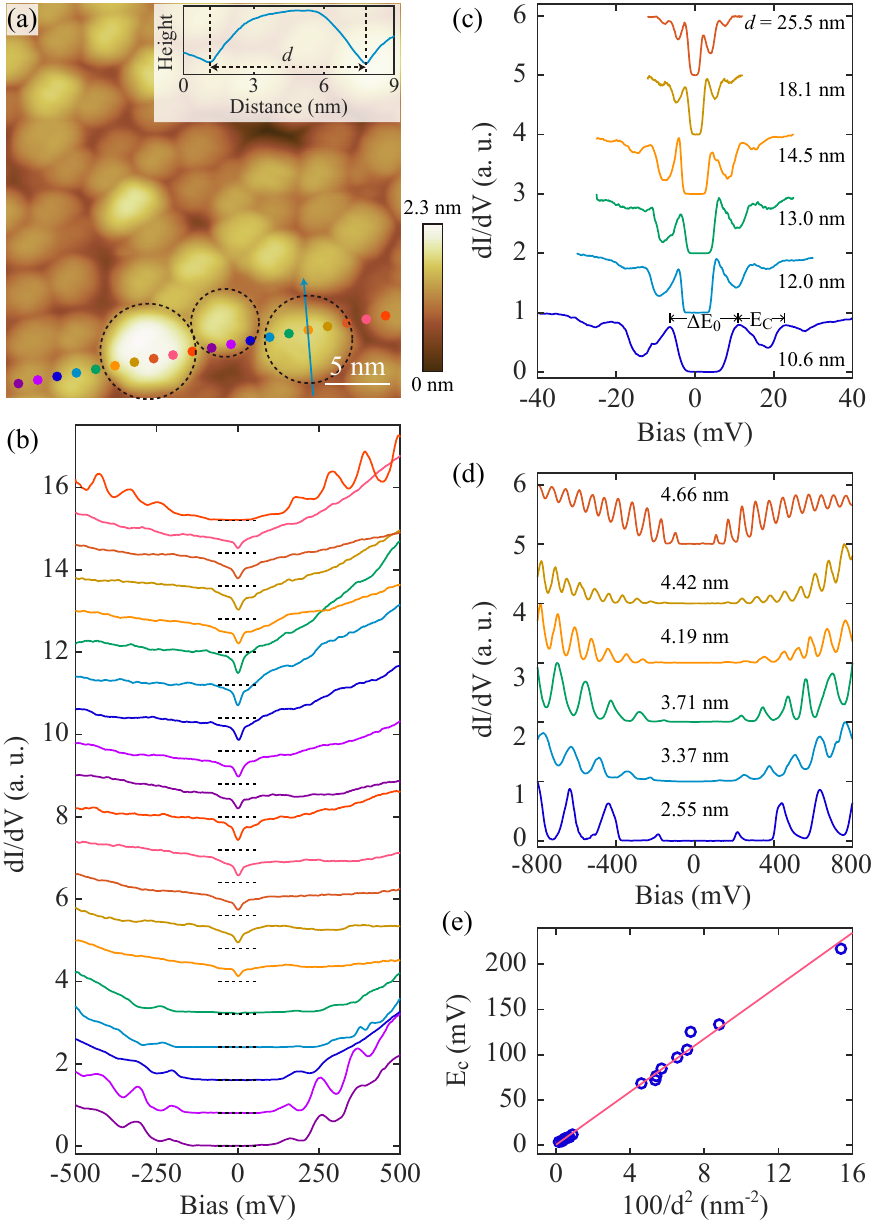}
\caption{\label{fig:2}(a) STM topographic image of PNSO nanoclusters (30 nm $\times$ 30 nm, $V$ = -3.0 V, $I$ = 10 pA). The nanoclusters are marked by dashed circles with various effective diameters ($d$). Inset: line profile along the blue arrow to determine the $d$ value. (b) A series of d$I$/d$V$ spectra (set points: $V$ = -500 mV, $I$ = 100 pA) acquired on the correspondingly colored spots in (a). (c), (d) Representative spectra on the nanoclusters of different sizes, as labeled. The set points in (c) are $V$ = 40 mV, 30 mV, 25 mV, 25 mV, 12 mV, 12 mV, $I$ = 1.5 nA, 8.0 nA, 6.0 nA, 6.0 nA, 1.3 nA, 2.0 nA for the spectra from bottom up. The set points in (d) are $V$ = -800 mV, $I$ = 0.6 nA for the 2.55-nm nanocluster and $V$ = -800 mV, $I$ = 1.0 nA for other sized nanoclusters. (e) Relationship between Coulomb staircase periodicity ($E_C$) and size parameters $1/d^2$. The red curve shows the best linear fit between them.}
\end{figure}

The PSNO nanoclusters are clearly imaged by a magnified STM image shown in Fig.~\ref{fig:2}(a). They are shaped as irregular polygons and mostly approximated as hemispheres. This allows us to measure their sizes by the diameter (labeled as $d$) of the outmost periphery of various nanoclusters. Figure~\ref{fig:2}(b) shows spatially resolved d$I$/d$V$ spectra straddling various nanoclusters taken at 4.2 K. These spectra exhibit a remarkable size $d$ dependence and can be divided into two categories. One is on the small nanoclusters that exhibits an energy gap around the Fermi level ($E_F$) and additional oscillatory peaks outside the low-energy gap. The other one is characteristic of a metal-like feature with an obvious spectral dip at $E_F$. To clarify the relationship between the spectra and the size of PSNO nanoclusters, we systematically measured the d$I$/d$V$ spectra as well as effective diameters for various nanoclusters, as illustrated in Figs.~\ref{fig:2}(c) and \ref{fig:2}(d). In general, all the spectra exhibit equidistant oscillatory peaks with increasing bias, reminiscent of Coulomb staircases for the single electron tunneling. According to this scenario, each peak means that the number of electrons in the nanoclusters changes by 1. As described in the DBTJ model \cite{hanna1991variation, averin1991theory, amman1991analytic}, the barrier increases with the decrease of the nanocluster size via $E_C = e/C_2$, where $C_2$ is the effective capacitance between the nanocluster and the continuous film beneath it. Figure~\ref{fig:2}(e) shows the dependence of $E_C$ on $d$ for various nanoclusters, justifying an inversely quadratic relationship between them via $E_C \propto 1/d^2$, because the capacitance of a metal island is proportional to the projected area in scale with $d^2$. Therefore, this result provides strong evidence for the notion that the equally spaced conductance peaks originate from Coulomb blockade effects in the PSNO nanoclusters.

\begin{figure*}
\includegraphics[width=1.5\columnwidth]{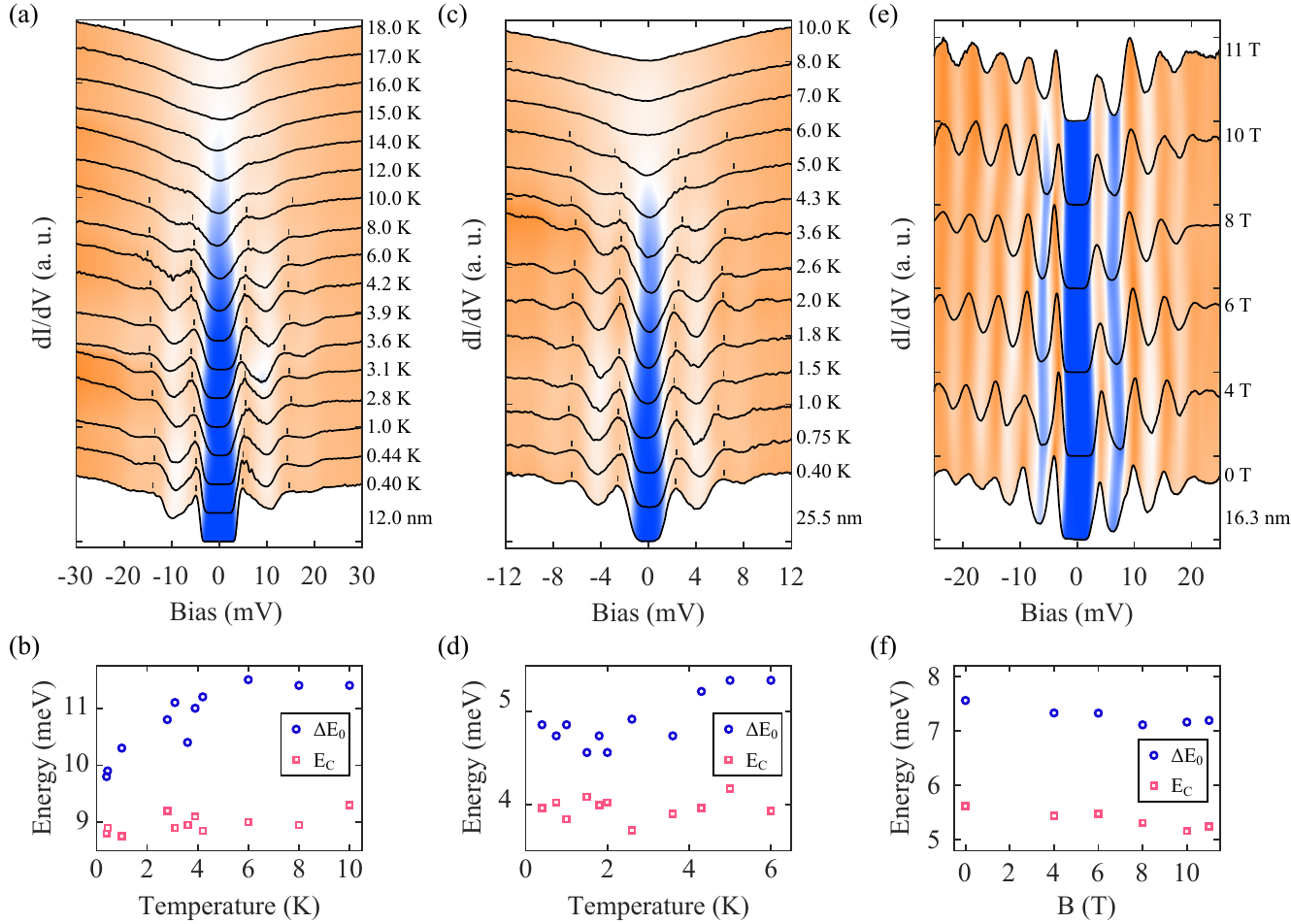}
\caption{\label{fig:3}(a) Temperature dependent d$I$/d$V$ spectra (set points: $V$ = 30 mV, $I$ = 8.0 nA) measured in a 12.0-nm PSNO nanocluster. (b) The extract ${\Delta}E_0$ and $E_C$ of spectra in (a) as a function of temperatures. (c), (d) Same as (a), (b) but on a 25.5-nm PSNO nanocluster. The set points are $V$ = 12 mV, $I$ = 2.0 nA. (e) A series of d$I$/d$V$ spectra (set points: $V$ = 25 mV, $I$ = 4.0 nA) measured in a 16.3-nm PSNO nanocluster under different magnetic fields. (f) The extract ${\Delta}E_0$ and $E_C$ of spectra in (e) as a function of magnetic fields.}
\end{figure*}

Note that the equidistant conductance peaks of the tunneling d$I$/d$V$ spectra hold true only for those well above and below $E_F$, whereas the energy spacing between the two peaks on both sides of $E_F$ (labeled as ${\Delta}E_0$) is commonly larger than $E_C$. A straightforward explanation may be that the gap ${\Delta}E_0$ contains the superconducting gap ($\Delta_S$) in PSNO. For most size-confined superconducting systems, the relationship ${\Delta}E_0 = E_C + 2\Delta_S$ holds when the cluster size exceeds the Anderson limit \cite{qin2020coupling, yuan2020observation, anderson1959theory}. At elevated temperature or magnetic field, the superconducting gap is suppressed and the gap ${\Delta}E_0$ will reduce from $E_C + 2\Delta_S$ to $E_C$, which provides a feasible criterion to distinguish superconductivity from the coupled Coulomb effect. To this purpose, temperature-dependent spectral investigations have been carried out on two typical nanoclusters with $d = $12.0 nm and $d = $25.5 nm [Figs.~\ref{fig:3}(a)-\ref{fig:3}(d)], respectively.  The extracted ${\Delta}E_0$ and $E_C$ from these spectra are shown in Figs.~\ref{fig:3}(b) and \ref{fig:3}(d), where the $E_C$ are extracted from the average of the energy differences between the first and second oscillatory peaks (marked by short rods) on both energy sides. Surprisingly, the gap ${\Delta}E_0$ never approaches $E_C$ with temperature, and the thermal broadening effect even broadens the two $E_F$-near conductance peaks such that ${\Delta}E_0$ phenomenally increases with temperature. At elevated temperatures, the gap-like features and conductance peaks gradually weaken and eventually disappear at the same time. The closing temperature of ${\Delta}E_0$ (6 K) in the 25.5-nm nanocluster is less than that (15 K) in the 12.0-nm nanocluster, which seems to be contradictory to the quantum size effect of superconductors that generally present a weaker superconductivity in smaller-sized nanoclusters \cite{bose2009competing}. Furthermore, we have also measured a series of d$I$/d$V$ spectra on varying magnetic fields in Figs.~\ref{fig:3}(e) and \ref{fig:3}(f), from which the extracted ${\Delta}E_0$ and $E_C$  change little with the increasing field. These results compellingly indicate that ${\Delta}E_0$ completely originates from the Coulomb effect and gets smeared out at elevated temperatures. No clear signature of the superconducting gap is found from our careful spectral measurements on the nano-clustered surface, even though the bulk superconductivity of the samples has been confirmed by transport measurements. Our results suggest that the nano-clustered surface has different stoichiometry from the bulk of the PSNO superconductors, where more caution should be paid in interpreting the gap-like features in the nickelate system.  

\begin{figure*}
\includegraphics[width=1.5\columnwidth]{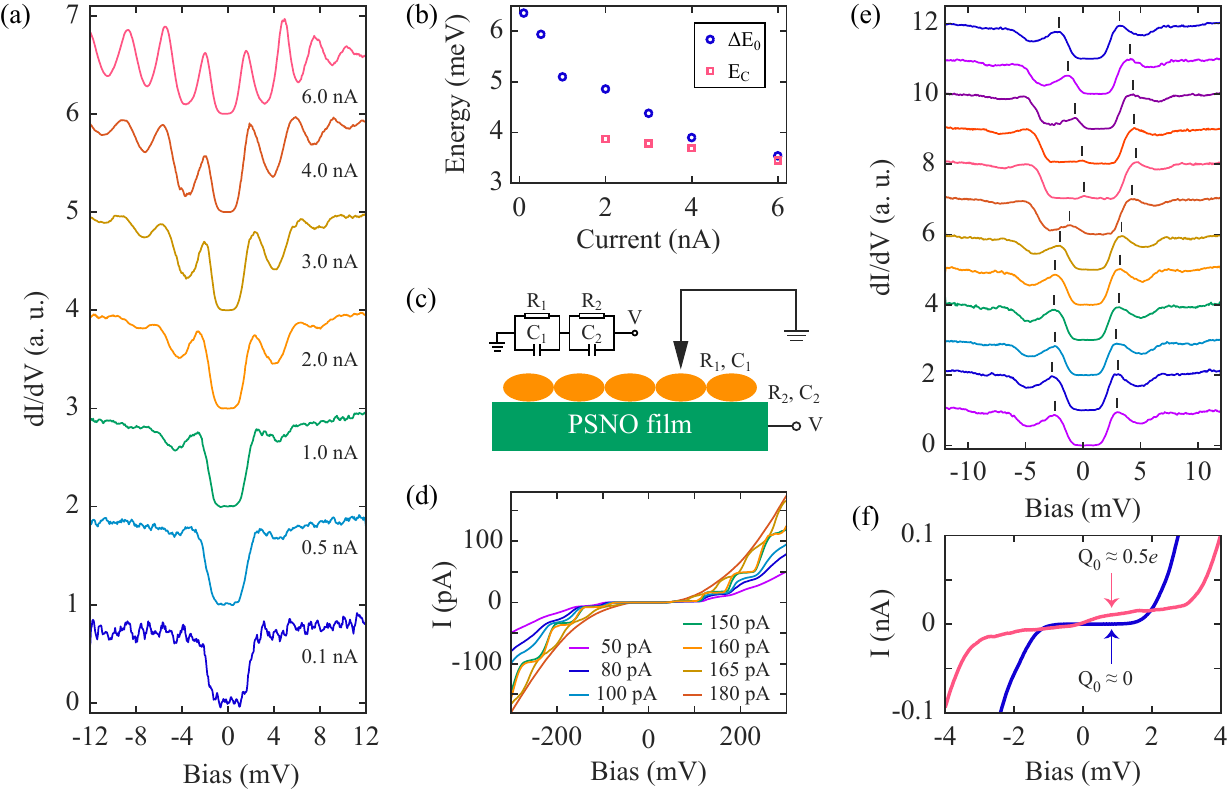}
\caption{\label{fig:4}(a) A series of d$I$/d$V$ spectra on a 25.5-nm PSNO nanocluster as a function of tunneling current set points. The bias set points are stabilized at 12 mV. (b) The extract ${\Delta}E_0$ and $E_C$ as a function of the current set points. (c) Schematic illustration of the DBTJ model on the PSNO surface and equivalent circuit of this structure. (d) A series of $I$-$V$ curves measured on a 4.66-nm PSNO nanocluster at different tunneling current set points. The bias set points are fixed at 300 mV. (e) A series of d$I$/d$V$ spectra (set points: $V$ = 12 mV, $I$ = 1.3 nA) acquired along a trajectory of 8.0 nm in an 18.1-nm PSNO nanocluster. (f) Comparison of $I$-$V$ curves between $Q_0 \approx 0$ and $Q_0 \approx 0.5e$.}
\end{figure*}

Having excluded the involved superconductivity, the DCB is considered as an alternative cause for the widened ${\Delta}E_0$. Experimentally, the DCB induced conductance gaps appear at high environmental impedance and are critically dependent on the tip-sample distances \cite{brun2012dynamical, senkpiel2020dynamical}. To check this scenario, Fig.~\ref{fig:4}(a) shows a series of site-specific normalized tunneling spectra on a 25.5-nm nanocluster at varied current set points. Evidently, there exists only a gap ${\Delta}E_0$ in the d$I$/d$V$ spectra measured at $I \leq 0.5$ nA, while the Coulomb staircases become more and more remarkable and ${\Delta}E_0$ gradually decreases to $E_C$ at larger $I$ [Fig.~\ref{fig:4}(b)]. These results show a salient crossover from Coulomb blockade (a dominant single junction) to Coulomb staircases (double-barrier junctions). As illustrated in Fig.~\ref{fig:4}(c), the whole experimental setup can be simplified to two tunneling junctions, just as in the DBTJ model. The first junction is between the tip and the surface PSNO nanocluster with an effective resistance $R_1$ and capacitance $C_1$, and the other one is between the surface PSNO nanocluster and the underlying PSNO film with effective parameters $R_2$ and $C_2$. When the distance between the tip and the sample is large enough, where $R_1$ is much larger than $R_2$, the first tunneling channel is dominant. However, as the tip approaches the sample surface so that $R_1$ is comparable with $R_2$, the second tunneling channel comes into play and the double junctions jointly cause the equidistant oscillatory peaks with the periodicity $E_C = e/C_2$. In special conditions (e.g. $I$ = 6.0 nA), the $I$-$V$ spectrum exhibits equally spaced staircases and the d$I$/d$V$ spectrum exhibits discrete peaks, where the influence of DCB is negligible. Figure~\ref{fig:4}(d) shows another series of measurements on a 4.66-nm nanocluster with larger $R_2$. One can immediately notice that the similar Coulomb staircases happen only for intermediate tunneling current set points (80 pA $\leq I \leq$ 165 pA). Too small or too large current set point $I$ means $R_1 \gg R_2$ or $R_1 \ll R_2$, which changes the configuration to a single junction model dominated by the first or second tunneling junction.

In addition to the effective resistances and capacitances, the fractional residual charge $Q_0$ is another important parameter in the DBTJ model, which represents the effective initial charge in the central electrode. Figure~\ref{fig:4}(e) illustrates the influence of $Q_0$ via a series of Coulomb staircases along a trajectory of 8.0 nm in an 18.1-nm nanocluster. These spectra are not symmetric with respect to $E_F$ but are shifted in energy because of the local variation of $Q_0$. In principle, the locations of the first spectral peaks below and above $E_F$ have analytical expressions $x_1 = (-e/2 + Q_0)/C_2$ and $x_2 = (e/2 + Q_0)/C_2$. As $Q_0$ approaches $e/2$, $x_1$ moves to the $E_F$ with minimum intensity. Figure~\ref{fig:4}(f) exhibits magnified $I$-$V$ spectra between $Q_0 \approx 0$ and $Q_0 \approx e/2$. At $Q_0 \approx 0$, the spectrum is characteristic of a Coulomb gap, while at $Q_0 \approx e/2$, the gap vanished and the $I$-$V$ curve shows a nonzero slope everywhere. The residual charge is considered to originate from the site-varying work functions and capacitances of the junctions \cite{hanna1991variation}. Given the almost identical shape of spectra beyond the first peaks, the capacitances should remain unchanged when the tip moves on the same nanocluster. Therefore, we ascribe the varying $Q_0$ to the variations of local work functions.

\section{SUMMARY}
In conclusion, the Coulomb blockade and staircases have been clearly identified on the clean surface of superconducting PSNO thin films after Ar$^+$ ion sputtering and annealing treatments. The remarkable Coulomb staircases are investigated as a function of the nanocluster sizes, tip-sample distances, and residual charges, which corroborate a SET process described by the DBTJ model. More importantly, the crossover from Coulomb blockade to Coulomb staircases has been well evidenced. Our results reveal the significant Coulomb blockade effects that should be carefully considered to explain the tunneling spectra of nano-clustered surface in infinite-layer nickelates.

\begin{acknowledgments}
The work at China was supported by the National Key R\&D of China (Grant No. 2022YFA1403100), the National Natural Science Foundation of China (Grant No. 12134008). The work at Stanford/SLAC was supported by the US Department of Energy, Office of Basic Energy Sciences, Division of Materials Sciences and Engineering, under contract no. DE-AC02-76SF00515, and Gordon and Betty Moore Foundation’s Emergent Phenomena in Quantum Systems Initiative through grant no. GBMF9072 (synthesis equipment). D.L. acknowledges the support from Hong Kong Research Grants Council (RGC) through ECS (CityU 21301221) and GRF (CityU 11309622) grants, and from National Science Foundation of China (NSFC12174325).
\end{acknowledgments}

\appendix*

\begin{figure}
\includegraphics{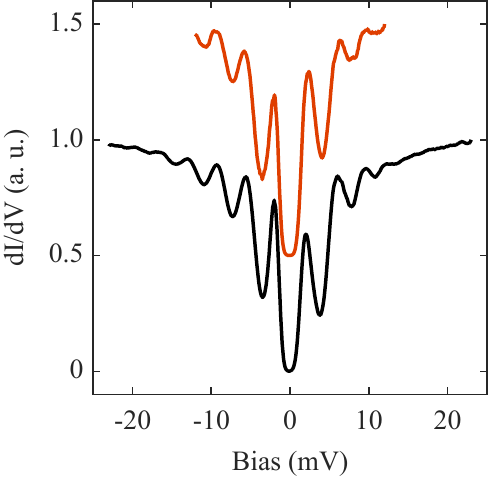}
\caption{\label{fig:5}Two d$I$/d$V$ spectra measured on the same nanocluster with different energy ranges. The spectra are vertically shifted for clarity. The set points of the lower spectrum are $V$ = 23 mV, $I$ = 8.0 nA and those of the upper spectrum are $V$ = 12 mV, $I$ = 3.0 nA.}
\end{figure}

\section{Spectra at different energy scales}
The oscillatory peaks from Coulomb staircases are obvious only when the effective resistances $R_1$ and $R_2$ satisfy an appropriate relationship, which makes us measure the d$I$/d$V$ spectra at different set-point conditions. However, it should be emphasized that the period $E_c$ does not change at a wide range of set points as long as the DBTJ forms [Figs.~\ref{fig:4}(a) and \ref{fig:4}(d)]. To further validate this point, we show two d$I$/d$V$ spectra at different energy scales in Fig.~\ref{fig:5}, where the spectrum at larger energy encompasses the oscillatory patterns of the one at smaller energy.

\bibliography{PSNO}

\begin{thebibliography}{37}%
\makeatletter
\providecommand \@ifxundefined [1]{%
 \@ifx{#1\undefined}
}%
\providecommand \@ifnum [1]{%
 \ifnum #1\expandafter \@firstoftwo
 \else \expandafter \@secondoftwo
 \fi
}%
\providecommand \@ifx [1]{%
 \ifx #1\expandafter \@firstoftwo
 \else \expandafter \@secondoftwo
 \fi
}%
\providecommand \natexlab [1]{#1}%
\providecommand \enquote  [1]{``#1''}%
\providecommand \bibnamefont  [1]{#1}%
\providecommand \bibfnamefont [1]{#1}%
\providecommand \citenamefont [1]{#1}%
\providecommand \href@noop [0]{\@secondoftwo}%
\providecommand \href [0]{\begingroup \@sanitize@url \@href}%
\providecommand \@href[1]{\@@startlink{#1}\@@href}%
\providecommand \@@href[1]{\endgroup#1\@@endlink}%
\providecommand \@sanitize@url [0]{\catcode `\\12\catcode `\$12\catcode
  `\&12\catcode `\#12\catcode `\^12\catcode `\_12\catcode `\%12\relax}%
\providecommand \@@startlink[1]{}%
\providecommand \@@endlink[0]{}%
\providecommand \url  [0]{\begingroup\@sanitize@url \@url }%
\providecommand \@url [1]{\endgroup\@href {#1}{\urlprefix }}%
\providecommand \urlprefix  [0]{URL }%
\providecommand \Eprint [0]{\href }%
\providecommand \doibase [0]{http://dx.doi.org/}%
\providecommand \selectlanguage [0]{\@gobble}%
\providecommand \bibinfo  [0]{\@secondoftwo}%
\providecommand \bibfield  [0]{\@secondoftwo}%
\providecommand \translation [1]{[#1]}%
\providecommand \BibitemOpen [0]{}%
\providecommand \bibitemStop [0]{}%
\providecommand \bibitemNoStop [0]{.\EOS\space}%
\providecommand \EOS [0]{\spacefactor3000\relax}%
\providecommand \BibitemShut  [1]{\csname bibitem#1\endcsname}%
\let\auto@bib@innerbib\@empty
\bibitem [{\citenamefont {Li}\ \emph {et~al.}(2019)\citenamefont {Li},
  \citenamefont {Lee}, \citenamefont {Wang}, \citenamefont {Osada},
  \citenamefont {Crossley}, \citenamefont {Lee}, \citenamefont {Cui},
  \citenamefont {Hikita},\ and\ \citenamefont
  {Hwang}}]{li2019superconductivity}%
  \BibitemOpen
  \bibfield  {author} {\bibinfo {author} {\bibfnamefont {Danfeng}\ \bibnamefont
  {Li}}, \bibinfo {author} {\bibfnamefont {Kyuho}\ \bibnamefont {Lee}},
  \bibinfo {author} {\bibfnamefont {Bai~Yang}\ \bibnamefont {Wang}}, \bibinfo
  {author} {\bibfnamefont {Motoki}\ \bibnamefont {Osada}}, \bibinfo {author}
  {\bibfnamefont {Samuel}\ \bibnamefont {Crossley}}, \bibinfo {author}
  {\bibfnamefont {Hye~Ryoung}\ \bibnamefont {Lee}}, \bibinfo {author}
  {\bibfnamefont {Yi}~\bibnamefont {Cui}}, \bibinfo {author} {\bibfnamefont
  {Yasuyuki}\ \bibnamefont {Hikita}}, \ and\ \bibinfo {author} {\bibfnamefont
  {Harold~Y}\ \bibnamefont {Hwang}},\ }\bibfield  {title} {\enquote {\bibinfo
  {title} {Superconductivity in an infinite-layer nickelate},}\ }\href
  {\doibase 10.1038/s41586-019-1496-5} {\bibfield  {journal} {\bibinfo
  {journal} {Nature}\ }\textbf {\bibinfo {volume} {572}},\ \bibinfo {pages}
  {624--627} (\bibinfo {year} {2019})}\BibitemShut {NoStop}%
\bibitem [{\citenamefont {Osada}\ \emph
  {et~al.}(2020{\natexlab{a}})\citenamefont {Osada}, \citenamefont {Wang},
  \citenamefont {Goodge}, \citenamefont {Lee}, \citenamefont {Yoon},
  \citenamefont {Sakuma}, \citenamefont {Li}, \citenamefont {Miura},
  \citenamefont {Kourkoutis},\ and\ \citenamefont
  {Hwang}}]{osada2020superconducting}%
  \BibitemOpen
  \bibfield  {author} {\bibinfo {author} {\bibfnamefont {Motoki}\ \bibnamefont
  {Osada}}, \bibinfo {author} {\bibfnamefont {Bai~Yang}\ \bibnamefont {Wang}},
  \bibinfo {author} {\bibfnamefont {Berit~H}\ \bibnamefont {Goodge}}, \bibinfo
  {author} {\bibfnamefont {Kyuho}\ \bibnamefont {Lee}}, \bibinfo {author}
  {\bibfnamefont {Hyeok}\ \bibnamefont {Yoon}}, \bibinfo {author}
  {\bibfnamefont {Keita}\ \bibnamefont {Sakuma}}, \bibinfo {author}
  {\bibfnamefont {Danfeng}\ \bibnamefont {Li}}, \bibinfo {author}
  {\bibfnamefont {Masashi}\ \bibnamefont {Miura}}, \bibinfo {author}
  {\bibfnamefont {Lena~F}\ \bibnamefont {Kourkoutis}}, \ and\ \bibinfo {author}
  {\bibfnamefont {Harold~Y}\ \bibnamefont {Hwang}},\ }\bibfield  {title}
  {\enquote {\bibinfo {title} {A superconducting praseodymium nickelate with
  infinite layer structure},}\ }\href {\doibase 10.1021/acs.nanolett.0c01392}
  {\bibfield  {journal} {\bibinfo  {journal} {Nano Lett.}\ }\textbf {\bibinfo
  {volume} {20}},\ \bibinfo {pages} {5735--5740} (\bibinfo {year}
  {2020}{\natexlab{a}})}\BibitemShut {NoStop}%
\bibitem [{\citenamefont {Zeng}\ \emph {et~al.}(2022)\citenamefont {Zeng},
  \citenamefont {Li}, \citenamefont {Chow}, \citenamefont {Cao}, \citenamefont
  {Zhang}, \citenamefont {Tang}, \citenamefont {Yin}, \citenamefont {Lim},
  \citenamefont {Hu}, \citenamefont {Yang} \emph
  {et~al.}}]{zeng2022superconductivity}%
  \BibitemOpen
  \bibfield  {author} {\bibinfo {author} {\bibfnamefont {Shengwei}\
  \bibnamefont {Zeng}}, \bibinfo {author} {\bibfnamefont {Changjian}\
  \bibnamefont {Li}}, \bibinfo {author} {\bibfnamefont {Lin~Er}\ \bibnamefont
  {Chow}}, \bibinfo {author} {\bibfnamefont {Yu}~\bibnamefont {Cao}}, \bibinfo
  {author} {\bibfnamefont {Zhaoting}\ \bibnamefont {Zhang}}, \bibinfo {author}
  {\bibfnamefont {Chi~Sin}\ \bibnamefont {Tang}}, \bibinfo {author}
  {\bibfnamefont {Xinmao}\ \bibnamefont {Yin}}, \bibinfo {author}
  {\bibfnamefont {Zhi~Shiuh}\ \bibnamefont {Lim}}, \bibinfo {author}
  {\bibfnamefont {Junxiong}\ \bibnamefont {Hu}}, \bibinfo {author}
  {\bibfnamefont {Ping}\ \bibnamefont {Yang}},  \emph {et~al.},\ }\bibfield
  {title} {\enquote {\bibinfo {title} {{Superconductivity in infinite-layer
  nickelate La$_{1-x}$Ca$_x$NiO$_2$ thin films}},}\ }\href {\doibase
  10.1126/sciadv.abl9927} {\bibfield  {journal} {\bibinfo  {journal} {Sci.
  Adv.}\ }\textbf {\bibinfo {volume} {8}},\ \bibinfo {pages} {eabl9927}
  (\bibinfo {year} {2022})}\BibitemShut {NoStop}%
\bibitem [{\citenamefont {Osada}\ \emph {et~al.}(2021)\citenamefont {Osada},
  \citenamefont {Wang}, \citenamefont {Goodge}, \citenamefont {Harvey},
  \citenamefont {Lee}, \citenamefont {Li}, \citenamefont {Kourkoutis},\ and\
  \citenamefont {Hwang}}]{osada2021nickelate}%
  \BibitemOpen
  \bibfield  {author} {\bibinfo {author} {\bibfnamefont {Motoki}\ \bibnamefont
  {Osada}}, \bibinfo {author} {\bibfnamefont {Bai~Yang}\ \bibnamefont {Wang}},
  \bibinfo {author} {\bibfnamefont {Berit~H}\ \bibnamefont {Goodge}}, \bibinfo
  {author} {\bibfnamefont {Shannon~P}\ \bibnamefont {Harvey}}, \bibinfo
  {author} {\bibfnamefont {Kyuho}\ \bibnamefont {Lee}}, \bibinfo {author}
  {\bibfnamefont {Danfeng}\ \bibnamefont {Li}}, \bibinfo {author}
  {\bibfnamefont {Lena~F}\ \bibnamefont {Kourkoutis}}, \ and\ \bibinfo {author}
  {\bibfnamefont {Harold~Y}\ \bibnamefont {Hwang}},\ }\bibfield  {title}
  {\enquote {\bibinfo {title} {{Nickelate superconductivity without rare-earth
  magnetism: (La, Sr)NiO$_2$}},}\ }\href {\doibase 10.1002/adma.202104083}
  {\bibfield  {journal} {\bibinfo  {journal} {Adv. Mater.}\ }\textbf {\bibinfo
  {volume} {33}},\ \bibinfo {pages} {2104083} (\bibinfo {year}
  {2021})}\BibitemShut {NoStop}%
\bibitem [{\citenamefont {Keimer}\ \emph {et~al.}(2015)\citenamefont {Keimer},
  \citenamefont {Kivelson}, \citenamefont {Norman}, \citenamefont {Uchida},\
  and\ \citenamefont {Zaanen}}]{keimer2015quantum}%
  \BibitemOpen
  \bibfield  {author} {\bibinfo {author} {\bibfnamefont {Bernhard}\
  \bibnamefont {Keimer}}, \bibinfo {author} {\bibfnamefont {Steven~A}\
  \bibnamefont {Kivelson}}, \bibinfo {author} {\bibfnamefont {Michael~R}\
  \bibnamefont {Norman}}, \bibinfo {author} {\bibfnamefont {Shinichi}\
  \bibnamefont {Uchida}}, \ and\ \bibinfo {author} {\bibfnamefont
  {J}~\bibnamefont {Zaanen}},\ }\bibfield  {title} {\enquote {\bibinfo {title}
  {From quantum matter to high-temperature superconductivity in copper
  oxides},}\ }\href {\doibase 10.1038/nature14165} {\bibfield  {journal}
  {\bibinfo  {journal} {Nature}\ }\textbf {\bibinfo {volume} {518}},\ \bibinfo
  {pages} {179--186} (\bibinfo {year} {2015})}\BibitemShut {NoStop}%
\bibitem [{\citenamefont {Li}\ \emph {et~al.}(2020)\citenamefont {Li},
  \citenamefont {Wang}, \citenamefont {Lee}, \citenamefont {Harvey},
  \citenamefont {Osada}, \citenamefont {Goodge}, \citenamefont {Kourkoutis},\
  and\ \citenamefont {Hwang}}]{li2020superconducting}%
  \BibitemOpen
  \bibfield  {author} {\bibinfo {author} {\bibfnamefont {Danfeng}\ \bibnamefont
  {Li}}, \bibinfo {author} {\bibfnamefont {Bai~Yang}\ \bibnamefont {Wang}},
  \bibinfo {author} {\bibfnamefont {Kyuho}\ \bibnamefont {Lee}}, \bibinfo
  {author} {\bibfnamefont {Shannon~P}\ \bibnamefont {Harvey}}, \bibinfo
  {author} {\bibfnamefont {Motoki}\ \bibnamefont {Osada}}, \bibinfo {author}
  {\bibfnamefont {Berit~H}\ \bibnamefont {Goodge}}, \bibinfo {author}
  {\bibfnamefont {Lena~F}\ \bibnamefont {Kourkoutis}}, \ and\ \bibinfo {author}
  {\bibfnamefont {Harold~Y}\ \bibnamefont {Hwang}},\ }\bibfield  {title}
  {\enquote {\bibinfo {title} {{Superconducting dome in Nd$_{1-x}$Sr$_x$NiO$_2$
  infinite layer films}},}\ }\href {\doibase 10.1103/PhysRevLett.125.027001}
  {\bibfield  {journal} {\bibinfo  {journal} {Phys. Rev. Lett.}\ }\textbf
  {\bibinfo {volume} {125}},\ \bibinfo {pages} {027001} (\bibinfo {year}
  {2020})}\BibitemShut {NoStop}%
\bibitem [{\citenamefont {Zeng}\ \emph {et~al.}(2020)\citenamefont {Zeng},
  \citenamefont {Tang}, \citenamefont {Yin}, \citenamefont {Li}, \citenamefont
  {Li}, \citenamefont {Huang}, \citenamefont {Hu}, \citenamefont {Liu},
  \citenamefont {Omar}, \citenamefont {Jani} \emph {et~al.}}]{zeng2020phase}%
  \BibitemOpen
  \bibfield  {author} {\bibinfo {author} {\bibfnamefont {Shengwei}\
  \bibnamefont {Zeng}}, \bibinfo {author} {\bibfnamefont {Chi~Sin}\
  \bibnamefont {Tang}}, \bibinfo {author} {\bibfnamefont {Xinmao}\ \bibnamefont
  {Yin}}, \bibinfo {author} {\bibfnamefont {Changjian}\ \bibnamefont {Li}},
  \bibinfo {author} {\bibfnamefont {Mengsha}\ \bibnamefont {Li}}, \bibinfo
  {author} {\bibfnamefont {Zhen}\ \bibnamefont {Huang}}, \bibinfo {author}
  {\bibfnamefont {Junxiong}\ \bibnamefont {Hu}}, \bibinfo {author}
  {\bibfnamefont {Wei}\ \bibnamefont {Liu}}, \bibinfo {author} {\bibfnamefont
  {Ganesh~Ji}\ \bibnamefont {Omar}}, \bibinfo {author} {\bibfnamefont {Hariom}\
  \bibnamefont {Jani}},  \emph {et~al.},\ }\bibfield  {title} {\enquote
  {\bibinfo {title} {{Phase diagram and superconducting dome of infinite-layer
  Nd$_{1-x}$Sr$_x$NiO$_2$ thin films}},}\ }\href {\doibase
  10.1103/PhysRevLett.125.147003} {\bibfield  {journal} {\bibinfo  {journal}
  {Phys. Rev. Lett.}\ }\textbf {\bibinfo {volume} {125}},\ \bibinfo {pages}
  {147003} (\bibinfo {year} {2020})}\BibitemShut {NoStop}%
\bibitem [{\citenamefont {Osada}\ \emph
  {et~al.}(2020{\natexlab{b}})\citenamefont {Osada}, \citenamefont {Wang},
  \citenamefont {Lee}, \citenamefont {Li},\ and\ \citenamefont
  {Hwang}}]{osada2020phase}%
  \BibitemOpen
  \bibfield  {author} {\bibinfo {author} {\bibfnamefont {Motoki}\ \bibnamefont
  {Osada}}, \bibinfo {author} {\bibfnamefont {Bai~Yang}\ \bibnamefont {Wang}},
  \bibinfo {author} {\bibfnamefont {Kyuho}\ \bibnamefont {Lee}}, \bibinfo
  {author} {\bibfnamefont {Danfeng}\ \bibnamefont {Li}}, \ and\ \bibinfo
  {author} {\bibfnamefont {Harold~Y}\ \bibnamefont {Hwang}},\ }\bibfield
  {title} {\enquote {\bibinfo {title} {{Phase diagram of infinite layer
  praseodymium nickelate Pr$_{1- x}$Sr$_x$NiO$_2$ thin films}},}\ }\href
  {\doibase 10.1103/PhysRevMaterials.4.121801} {\bibfield  {journal} {\bibinfo
  {journal} {Phys. Rev. Mater.}\ }\textbf {\bibinfo {volume} {4}},\ \bibinfo
  {pages} {121801} (\bibinfo {year} {2020}{\natexlab{b}})}\BibitemShut
  {NoStop}%
\bibitem [{\citenamefont {Lee}\ \emph {et~al.}(2022)\citenamefont {Lee},
  \citenamefont {Wang}, \citenamefont {Osada}, \citenamefont {Goodge},
  \citenamefont {Wang}, \citenamefont {Lee}, \citenamefont {Harvey},
  \citenamefont {Kim}, \citenamefont {Yu}, \citenamefont {Murthy} \emph
  {et~al.}}]{lee2022character}%
  \BibitemOpen
  \bibfield  {author} {\bibinfo {author} {\bibfnamefont {Kyuho}\ \bibnamefont
  {Lee}}, \bibinfo {author} {\bibfnamefont {Bai~Yang}\ \bibnamefont {Wang}},
  \bibinfo {author} {\bibfnamefont {Motoki}\ \bibnamefont {Osada}}, \bibinfo
  {author} {\bibfnamefont {Berit~H}\ \bibnamefont {Goodge}}, \bibinfo {author}
  {\bibfnamefont {Tiffany~C}\ \bibnamefont {Wang}}, \bibinfo {author}
  {\bibfnamefont {Yonghun}\ \bibnamefont {Lee}}, \bibinfo {author}
  {\bibfnamefont {Shannon}\ \bibnamefont {Harvey}}, \bibinfo {author}
  {\bibfnamefont {Woo~Jin}\ \bibnamefont {Kim}}, \bibinfo {author}
  {\bibfnamefont {Yijun}\ \bibnamefont {Yu}}, \bibinfo {author} {\bibfnamefont
  {Chaitanya}\ \bibnamefont {Murthy}},  \emph {et~al.},\ }\bibfield  {title}
  {\enquote {\bibinfo {title} {Character of the "normal state" of the nickelate
  superconductors},}\ }\href {\doibase 10.48550/arXiv.2203.02580} {\bibfield
  {journal} {\bibinfo  {journal} {arXiv preprint arXiv:2203.02580}\ } (\bibinfo
  {year} {2022}),\ 10.48550/arXiv.2203.02580}\BibitemShut {NoStop}%
\bibitem [{\citenamefont {Hu}\ and\ \citenamefont {Wu}(2019)}]{hu2019two}%
  \BibitemOpen
  \bibfield  {author} {\bibinfo {author} {\bibfnamefont {Lun-Hui}\ \bibnamefont
  {Hu}}\ and\ \bibinfo {author} {\bibfnamefont {Congjun}\ \bibnamefont {Wu}},\
  }\bibfield  {title} {\enquote {\bibinfo {title} {Two-band model for magnetism
  and superconductivity in nickelates},}\ }\href {\doibase
  10.1103/PhysRevResearch.1.032046} {\bibfield  {journal} {\bibinfo  {journal}
  {Phys. Rev. Res.}\ }\textbf {\bibinfo {volume} {1}},\ \bibinfo {pages}
  {032046} (\bibinfo {year} {2019})}\BibitemShut {NoStop}%
\bibitem [{\citenamefont {Botana}\ and\ \citenamefont
  {Norman}(2020)}]{botana2020similarities}%
  \BibitemOpen
  \bibfield  {author} {\bibinfo {author} {\bibfnamefont {Antia~S}\ \bibnamefont
  {Botana}}\ and\ \bibinfo {author} {\bibfnamefont {Michael~R}\ \bibnamefont
  {Norman}},\ }\bibfield  {title} {\enquote {\bibinfo {title} {{Similarities
  and differences between LaNiO$_2$ and CaCuO$_2$ and implications for
  superconductivity}},}\ }\href {\doibase 10.1103/PhysRevX.10.011024}
  {\bibfield  {journal} {\bibinfo  {journal} {Phys. Rev. X}\ }\textbf {\bibinfo
  {volume} {10}},\ \bibinfo {pages} {011024} (\bibinfo {year}
  {2020})}\BibitemShut {NoStop}%
\bibitem [{\citenamefont {Choi}\ \emph {et~al.}(2020)\citenamefont {Choi},
  \citenamefont {Lee},\ and\ \citenamefont {Pickett}}]{choi2020role}%
  \BibitemOpen
  \bibfield  {author} {\bibinfo {author} {\bibfnamefont {Mi-Young}\
  \bibnamefont {Choi}}, \bibinfo {author} {\bibfnamefont {Kwan-Woo}\
  \bibnamefont {Lee}}, \ and\ \bibinfo {author} {\bibfnamefont {Warren~E}\
  \bibnamefont {Pickett}},\ }\bibfield  {title} {\enquote {\bibinfo {title}
  {{Role of 4f states in infinite-layer NdNiO$_2$}},}\ }\href {\doibase
  10.1103/PhysRevB.101.020503} {\bibfield  {journal} {\bibinfo  {journal}
  {Phys. Rev. B}\ }\textbf {\bibinfo {volume} {101}},\ \bibinfo {pages}
  {020503} (\bibinfo {year} {2020})}\BibitemShut {NoStop}%
\bibitem [{\citenamefont {Fischer}\ \emph {et~al.}(2007)\citenamefont
  {Fischer}, \citenamefont {Kugler}, \citenamefont {Maggio-Aprile},
  \citenamefont {Berthod},\ and\ \citenamefont {Renner}}]{fischer2007scanning}%
  \BibitemOpen
  \bibfield  {author} {\bibinfo {author} {\bibfnamefont {{\O}ystein}\
  \bibnamefont {Fischer}}, \bibinfo {author} {\bibfnamefont {Martin}\
  \bibnamefont {Kugler}}, \bibinfo {author} {\bibfnamefont {Ivan}\ \bibnamefont
  {Maggio-Aprile}}, \bibinfo {author} {\bibfnamefont {Christophe}\ \bibnamefont
  {Berthod}}, \ and\ \bibinfo {author} {\bibfnamefont {Christoph}\ \bibnamefont
  {Renner}},\ }\bibfield  {title} {\enquote {\bibinfo {title} {Scanning
  tunneling spectroscopy of high-temperature superconductors},}\ }\href
  {\doibase 10.1103/RevModPhys.79.353} {\bibfield  {journal} {\bibinfo
  {journal} {Rev. Mod. Phys.}\ }\textbf {\bibinfo {volume} {79}},\ \bibinfo
  {pages} {353} (\bibinfo {year} {2007})}\BibitemShut {NoStop}%
\bibitem [{\citenamefont {Hoffman}(2011)}]{hoffman2011spectroscopic}%
  \BibitemOpen
  \bibfield  {author} {\bibinfo {author} {\bibfnamefont {Jennifer~E}\
  \bibnamefont {Hoffman}},\ }\bibfield  {title} {\enquote {\bibinfo {title}
  {{Spectroscopic scanning tunneling microscopy insights into Fe-based
  superconductors}},}\ }\href {\doibase 10.1088/0034-4885/74/12/124513}
  {\bibfield  {journal} {\bibinfo  {journal} {Rep. Prog. Phys.}\ }\textbf
  {\bibinfo {volume} {74}},\ \bibinfo {pages} {124513} (\bibinfo {year}
  {2011})}\BibitemShut {NoStop}%
\bibitem [{\citenamefont {Gu}\ \emph {et~al.}(2020)\citenamefont {Gu},
  \citenamefont {Li}, \citenamefont {Wan}, \citenamefont {Li}, \citenamefont
  {Guo}, \citenamefont {Yang}, \citenamefont {Li}, \citenamefont {Zhu},
  \citenamefont {Pan}, \citenamefont {Nie},\ and\ \citenamefont
  {Wen}}]{gu2020single}%
  \BibitemOpen
  \bibfield  {author} {\bibinfo {author} {\bibfnamefont {Qiangqiang}\
  \bibnamefont {Gu}}, \bibinfo {author} {\bibfnamefont {Yueying}\ \bibnamefont
  {Li}}, \bibinfo {author} {\bibfnamefont {Siyuan}\ \bibnamefont {Wan}},
  \bibinfo {author} {\bibfnamefont {Huazhou}\ \bibnamefont {Li}}, \bibinfo
  {author} {\bibfnamefont {Wei}\ \bibnamefont {Guo}}, \bibinfo {author}
  {\bibfnamefont {Huan}\ \bibnamefont {Yang}}, \bibinfo {author} {\bibfnamefont
  {Qing}\ \bibnamefont {Li}}, \bibinfo {author} {\bibfnamefont {Xiyu}\
  \bibnamefont {Zhu}}, \bibinfo {author} {\bibfnamefont {Xiaoqing}\
  \bibnamefont {Pan}}, \bibinfo {author} {\bibfnamefont {Yuefeng}\ \bibnamefont
  {Nie}}, \ and\ \bibinfo {author} {\bibfnamefont {Hai-Hu}\ \bibnamefont
  {Wen}},\ }\bibfield  {title} {\enquote {\bibinfo {title} {{Single particle
  tunneling spectrum of superconducting Nd$_{1-x}$Sr$_x$NiO$_2$ thin films}},}\
  }\href {\doibase 10.1038/s41467-020-19908-1} {\bibfield  {journal} {\bibinfo
  {journal} {Nat. Commun.}\ }\textbf {\bibinfo {volume} {11}},\ \bibinfo
  {pages} {1--7} (\bibinfo {year} {2020})}\BibitemShut {NoStop}%
\bibitem [{\citenamefont {Gu}\ and\ \citenamefont
  {Wen}(2021)}]{gu2021superconductivity}%
  \BibitemOpen
  \bibfield  {author} {\bibinfo {author} {\bibfnamefont {Qiangqiang}\
  \bibnamefont {Gu}}\ and\ \bibinfo {author} {\bibfnamefont {Haihu}\
  \bibnamefont {Wen}},\ }\bibfield  {title} {\enquote {\bibinfo {title}
  {Superconductivity in nickel based 112 systems},}\ }\href {\doibase
  10.1016/j.xinn.2021.100202} {\bibfield  {journal} {\bibinfo  {journal} {The
  Innovation}\ }\textbf {\bibinfo {volume} {3}},\ \bibinfo {pages} {100202}
  (\bibinfo {year} {2021})}\BibitemShut {NoStop}%
\bibitem [{\citenamefont {Adhikary}\ \emph {et~al.}(2020)\citenamefont
  {Adhikary}, \citenamefont {Bandyopadhyay}, \citenamefont {Das}, \citenamefont
  {Dasgupta},\ and\ \citenamefont {Saha-Dasgupta}}]{adhikary2020orbital}%
  \BibitemOpen
  \bibfield  {author} {\bibinfo {author} {\bibfnamefont {Priyo}\ \bibnamefont
  {Adhikary}}, \bibinfo {author} {\bibfnamefont {Subhadeep}\ \bibnamefont
  {Bandyopadhyay}}, \bibinfo {author} {\bibfnamefont {Tanmoy}\ \bibnamefont
  {Das}}, \bibinfo {author} {\bibfnamefont {Indra}\ \bibnamefont {Dasgupta}}, \
  and\ \bibinfo {author} {\bibfnamefont {Tanusri}\ \bibnamefont
  {Saha-Dasgupta}},\ }\bibfield  {title} {\enquote {\bibinfo {title}
  {Orbital-selective superconductivity in a two-band model of infinite-layer
  nickelates},}\ }\href {\doibase 10.1103/PhysRevB.102.100501} {\bibfield
  {journal} {\bibinfo  {journal} {Phys. Rev. B}\ }\textbf {\bibinfo {volume}
  {102}},\ \bibinfo {pages} {100501} (\bibinfo {year} {2020})}\BibitemShut
  {NoStop}%
\bibitem [{\citenamefont {Wang}\ \emph {et~al.}(2020)\citenamefont {Wang},
  \citenamefont {Zhang}, \citenamefont {Yang},\ and\ \citenamefont
  {Zhang}}]{wang2020distinct}%
  \BibitemOpen
  \bibfield  {author} {\bibinfo {author} {\bibfnamefont {Zhan}\ \bibnamefont
  {Wang}}, \bibinfo {author} {\bibfnamefont {Guang-Ming}\ \bibnamefont
  {Zhang}}, \bibinfo {author} {\bibfnamefont {Yi-feng}\ \bibnamefont {Yang}}, \
  and\ \bibinfo {author} {\bibfnamefont {Fu-Chun}\ \bibnamefont {Zhang}},\
  }\bibfield  {title} {\enquote {\bibinfo {title} {Distinct pairing symmetries
  of superconductivity in infinite-layer nickelates},}\ }\href {\doibase
  10.1103/PhysRevB.102.220501} {\bibfield  {journal} {\bibinfo  {journal}
  {Phys. Rev. B}\ }\textbf {\bibinfo {volume} {102}},\ \bibinfo {pages}
  {220501} (\bibinfo {year} {2020})}\BibitemShut {NoStop}%
\bibitem [{\citenamefont {Wu}\ \emph {et~al.}(2020)\citenamefont {Wu},
  \citenamefont {Jiang}, \citenamefont {Di~Sante}, \citenamefont {Hanke},
  \citenamefont {Schnyder}, \citenamefont {Hu},\ and\ \citenamefont
  {Thomale}}]{wu2020surface}%
  \BibitemOpen
  \bibfield  {author} {\bibinfo {author} {\bibfnamefont {Xianxin}\ \bibnamefont
  {Wu}}, \bibinfo {author} {\bibfnamefont {Kun}\ \bibnamefont {Jiang}},
  \bibinfo {author} {\bibfnamefont {Domenico}\ \bibnamefont {Di~Sante}},
  \bibinfo {author} {\bibfnamefont {Werner}\ \bibnamefont {Hanke}}, \bibinfo
  {author} {\bibfnamefont {AP}~\bibnamefont {Schnyder}}, \bibinfo {author}
  {\bibfnamefont {Jiangping}\ \bibnamefont {Hu}}, \ and\ \bibinfo {author}
  {\bibfnamefont {Ronny}\ \bibnamefont {Thomale}},\ }\bibfield  {title}
  {\enquote {\bibinfo {title} {Surface $s$-wave superconductivity for
  oxide-terminated infinite-layer nickelates},}\ }\href {\doibase
  10.48550/arXiv.2008.06009} {\bibfield  {journal} {\bibinfo  {journal} {arXiv
  preprint arXiv:2008.06009}\ } (\bibinfo {year} {2020}),\
  10.48550/arXiv.2008.06009}\BibitemShut {NoStop}%
\bibitem [{\citenamefont {Choubey}\ and\ \citenamefont
  {Eremin}(2021)}]{choubey2021electronic}%
  \BibitemOpen
  \bibfield  {author} {\bibinfo {author} {\bibfnamefont {Peayush}\ \bibnamefont
  {Choubey}}\ and\ \bibinfo {author} {\bibfnamefont {Ilya~M}\ \bibnamefont
  {Eremin}},\ }\bibfield  {title} {\enquote {\bibinfo {title} {Electronic
  theory for scanning tunneling microscopy spectra in infinite-layer nickelate
  superconductors},}\ }\href {\doibase 10.1103/PhysRevB.104.144504} {\bibfield
  {journal} {\bibinfo  {journal} {Phys. Rev. B}\ }\textbf {\bibinfo {volume}
  {104}},\ \bibinfo {pages} {144504} (\bibinfo {year} {2021})}\BibitemShut
  {NoStop}%
\bibitem [{\citenamefont {Perenboom}\ \emph {et~al.}(1981)\citenamefont
  {Perenboom}, \citenamefont {Wyder},\ and\ \citenamefont
  {Meier}}]{perenboom1981electronic}%
  \BibitemOpen
  \bibfield  {author} {\bibinfo {author} {\bibfnamefont {Johannes Antonius
  Albertus~Joseph}\ \bibnamefont {Perenboom}}, \bibinfo {author} {\bibfnamefont
  {Peter}\ \bibnamefont {Wyder}}, \ and\ \bibinfo {author} {\bibfnamefont
  {Felix}\ \bibnamefont {Meier}},\ }\bibfield  {title} {\enquote {\bibinfo
  {title} {Electronic properties of small metallic particles},}\ }\href
  {\doibase 10.1016/0370-1573(81)90194-0} {\bibfield  {journal} {\bibinfo
  {journal} {Phys. Rep.}\ }\textbf {\bibinfo {volume} {78}},\ \bibinfo {pages}
  {173--292} (\bibinfo {year} {1981})}\BibitemShut {NoStop}%
\bibitem [{\citenamefont {Likharev}(1988)}]{likharev1988correlated}%
  \BibitemOpen
  \bibfield  {author} {\bibinfo {author} {\bibfnamefont {Konstantin~K}\
  \bibnamefont {Likharev}},\ }\bibfield  {title} {\enquote {\bibinfo {title}
  {Correlated discrete transfer of single electrons in ultrasmall tunnel
  junctions},}\ }\href {\doibase 10.1147/rd.321.0144} {\bibfield  {journal}
  {\bibinfo  {journal} {IBM J. Res. Dev.}\ }\textbf {\bibinfo {volume} {32}},\
  \bibinfo {pages} {144--158} (\bibinfo {year} {1988})}\BibitemShut {NoStop}%
\bibitem [{\citenamefont {Zhang}\ \emph {et~al.}(2005)\citenamefont {Zhang},
  \citenamefont {Yasutake}, \citenamefont {Shichibu}, \citenamefont
  {Teranishi},\ and\ \citenamefont {Majima}}]{zhang2005tunneling}%
  \BibitemOpen
  \bibfield  {author} {\bibinfo {author} {\bibfnamefont {Hong}\ \bibnamefont
  {Zhang}}, \bibinfo {author} {\bibfnamefont {Yuhsuke}\ \bibnamefont
  {Yasutake}}, \bibinfo {author} {\bibfnamefont {Yuhkatsu}\ \bibnamefont
  {Shichibu}}, \bibinfo {author} {\bibfnamefont {Toshiharu}\ \bibnamefont
  {Teranishi}}, \ and\ \bibinfo {author} {\bibfnamefont {Yutaka}\ \bibnamefont
  {Majima}},\ }\bibfield  {title} {\enquote {\bibinfo {title} {{Tunneling
  resistance of double-barrier tunneling structures with an
  alkanethiol-protected Au nanoparticle}},}\ }\href {\doibase
  10.1103/PhysRevB.72.205441} {\bibfield  {journal} {\bibinfo  {journal} {Phys.
  Rev. B}\ }\textbf {\bibinfo {volume} {72}},\ \bibinfo {pages} {205441}
  (\bibinfo {year} {2005})}\BibitemShut {NoStop}%
\bibitem [{\citenamefont {Kano}\ \emph {et~al.}(2015)\citenamefont {Kano},
  \citenamefont {Tada},\ and\ \citenamefont {Majima}}]{kano2015nanoparticle}%
  \BibitemOpen
  \bibfield  {author} {\bibinfo {author} {\bibfnamefont {Shinya}\ \bibnamefont
  {Kano}}, \bibinfo {author} {\bibfnamefont {Tsukasa}\ \bibnamefont {Tada}}, \
  and\ \bibinfo {author} {\bibfnamefont {Yutaka}\ \bibnamefont {Majima}},\
  }\bibfield  {title} {\enquote {\bibinfo {title} {{Nanoparticle
  characterization based on STM and STS}},}\ }\href {\doibase
  10.1039/C4CS00204K} {\bibfield  {journal} {\bibinfo  {journal} {Chem. Soc.
  Rev.}\ }\textbf {\bibinfo {volume} {44}},\ \bibinfo {pages} {970--987}
  (\bibinfo {year} {2015})}\BibitemShut {NoStop}%
\bibitem [{\citenamefont {Sch{\"o}nenberger}\ \emph {et~al.}(1992)\citenamefont
  {Sch{\"o}nenberger}, \citenamefont {Van~Houten},\ and\ \citenamefont
  {Donkersloot}}]{schonenberger1992single}%
  \BibitemOpen
  \bibfield  {author} {\bibinfo {author} {\bibfnamefont {C}~\bibnamefont
  {Sch{\"o}nenberger}}, \bibinfo {author} {\bibfnamefont {H}~\bibnamefont
  {Van~Houten}}, \ and\ \bibinfo {author} {\bibfnamefont {HC}~\bibnamefont
  {Donkersloot}},\ }\bibfield  {title} {\enquote {\bibinfo {title}
  {Single-electron tunnelling observed at room temperature by
  scanning-tunnelling microscopy},}\ }\href {\doibase
  10.1209/0295-5075/20/3/010} {\bibfield  {journal} {\bibinfo  {journal}
  {Europhys. Lett.}\ }\textbf {\bibinfo {volume} {20}},\ \bibinfo {pages} {249}
  (\bibinfo {year} {1992})}\BibitemShut {NoStop}%
\bibitem [{\citenamefont {Dubois}\ \emph {et~al.}(1996)\citenamefont {Dubois},
  \citenamefont {Gerritsen}, \citenamefont {Shafranjuk}, \citenamefont {Boon},
  \citenamefont {Schmid},\ and\ \citenamefont
  {Van~Kempen}}]{dubois1996coulomb}%
  \BibitemOpen
  \bibfield  {author} {\bibinfo {author} {\bibfnamefont {JGA}\ \bibnamefont
  {Dubois}}, \bibinfo {author} {\bibfnamefont {JW}~\bibnamefont {Gerritsen}},
  \bibinfo {author} {\bibfnamefont {SE}~\bibnamefont {Shafranjuk}}, \bibinfo
  {author} {\bibfnamefont {EJG}\ \bibnamefont {Boon}}, \bibinfo {author}
  {\bibfnamefont {G}~\bibnamefont {Schmid}}, \ and\ \bibinfo {author}
  {\bibfnamefont {H}~\bibnamefont {Van~Kempen}},\ }\bibfield  {title} {\enquote
  {\bibinfo {title} {Coulomb staircases and quantum size effects in tunnelling
  spectroscopy on ligand-stabilized metal clusters},}\ }\href {\doibase
  10.1209/epl/i1996-00333-0} {\bibfield  {journal} {\bibinfo  {journal}
  {Europhys. Lett.}\ }\textbf {\bibinfo {volume} {33}},\ \bibinfo {pages} {279}
  (\bibinfo {year} {1996})}\BibitemShut {NoStop}%
\bibitem [{\citenamefont {Qin}\ \emph {et~al.}(2020)\citenamefont {Qin},
  \citenamefont {Zhao}, \citenamefont {Xia}, \citenamefont {Wang},
  \citenamefont {Liu}, \citenamefont {Guan}, \citenamefont {Wang},
  \citenamefont {Li}, \citenamefont {Zheng}, \citenamefont {Liu} \emph
  {et~al.}}]{qin2020coupling}%
  \BibitemOpen
  \bibfield  {author} {\bibinfo {author} {\bibfnamefont {Jin}\ \bibnamefont
  {Qin}}, \bibinfo {author} {\bibfnamefont {Chenxiao}\ \bibnamefont {Zhao}},
  \bibinfo {author} {\bibfnamefont {Bing}\ \bibnamefont {Xia}}, \bibinfo
  {author} {\bibfnamefont {Zerui}\ \bibnamefont {Wang}}, \bibinfo {author}
  {\bibfnamefont {Yu}~\bibnamefont {Liu}}, \bibinfo {author} {\bibfnamefont
  {Dandan}\ \bibnamefont {Guan}}, \bibinfo {author} {\bibfnamefont {Shiyong}\
  \bibnamefont {Wang}}, \bibinfo {author} {\bibfnamefont {Yaoyi}\ \bibnamefont
  {Li}}, \bibinfo {author} {\bibfnamefont {Hao}\ \bibnamefont {Zheng}},
  \bibinfo {author} {\bibfnamefont {Canhua}\ \bibnamefont {Liu}},  \emph
  {et~al.},\ }\bibfield  {title} {\enquote {\bibinfo {title} {{Coupling of
  superconductivity and Coulomb blockade in Sn nanoparticles}},}\ }\href
  {\doibase 10.1088/1361-6528/ab8763} {\bibfield  {journal} {\bibinfo
  {journal} {Nanotechnology}\ }\textbf {\bibinfo {volume} {31}},\ \bibinfo
  {pages} {305708} (\bibinfo {year} {2020})}\BibitemShut {NoStop}%
\bibitem [{\citenamefont {Yuan}\ \emph {et~al.}(2020)\citenamefont {Yuan},
  \citenamefont {Wang}, \citenamefont {Song}, \citenamefont {Wang},
  \citenamefont {He}, \citenamefont {Ma}, \citenamefont {Yao}, \citenamefont
  {Li},\ and\ \citenamefont {Xue}}]{yuan2020observation}%
  \BibitemOpen
  \bibfield  {author} {\bibinfo {author} {\bibfnamefont {Yonghao}\ \bibnamefont
  {Yuan}}, \bibinfo {author} {\bibfnamefont {Xintong}\ \bibnamefont {Wang}},
  \bibinfo {author} {\bibfnamefont {Canli}\ \bibnamefont {Song}}, \bibinfo
  {author} {\bibfnamefont {Lili}\ \bibnamefont {Wang}}, \bibinfo {author}
  {\bibfnamefont {Ke}~\bibnamefont {He}}, \bibinfo {author} {\bibfnamefont
  {Xucun}\ \bibnamefont {Ma}}, \bibinfo {author} {\bibfnamefont {Hong}\
  \bibnamefont {Yao}}, \bibinfo {author} {\bibfnamefont {Wei}\ \bibnamefont
  {Li}}, \ and\ \bibinfo {author} {\bibfnamefont {Qi-Kun}\ \bibnamefont
  {Xue}},\ }\bibfield  {title} {\enquote {\bibinfo {title} {{Observation of
  coulomb gap and enhanced superconducting gap in nano-sized Pb islands grown
  on SrTiO$_3$}},}\ }\href {\doibase 10.1088/0256-307X/37/1/017402} {\bibfield
  {journal} {\bibinfo  {journal} {Chin. Phys. Lett.}\ }\textbf {\bibinfo
  {volume} {37}},\ \bibinfo {pages} {017402} (\bibinfo {year}
  {2020})}\BibitemShut {NoStop}%
\bibitem [{\citenamefont {Hanna}\ and\ \citenamefont
  {Tinkham}(1991)}]{hanna1991variation}%
  \BibitemOpen
  \bibfield  {author} {\bibinfo {author} {\bibfnamefont {AE}~\bibnamefont
  {Hanna}}\ and\ \bibinfo {author} {\bibfnamefont {M}~\bibnamefont {Tinkham}},\
  }\bibfield  {title} {\enquote {\bibinfo {title} {{Variation of the Coulomb
  staircase in a two-junction system by fractional electron charge}},}\ }\href
  {\doibase 10.1103/PhysRevB.44.5919} {\bibfield  {journal} {\bibinfo
  {journal} {Phys. Rev. B}\ }\textbf {\bibinfo {volume} {44}},\ \bibinfo
  {pages} {5919} (\bibinfo {year} {1991})}\BibitemShut {NoStop}%
\bibitem [{\citenamefont {Averin}\ \emph {et~al.}(1991)\citenamefont {Averin},
  \citenamefont {Korotkov},\ and\ \citenamefont {Likharev}}]{averin1991theory}%
  \BibitemOpen
  \bibfield  {author} {\bibinfo {author} {\bibfnamefont {DV}~\bibnamefont
  {Averin}}, \bibinfo {author} {\bibfnamefont {AN}~\bibnamefont {Korotkov}}, \
  and\ \bibinfo {author} {\bibfnamefont {KK}~\bibnamefont {Likharev}},\
  }\bibfield  {title} {\enquote {\bibinfo {title} {Theory of single-electron
  charging of quantum wells and dots},}\ }\href {\doibase
  10.1103/PhysRevB.44.6199} {\bibfield  {journal} {\bibinfo  {journal} {Phys.
  Rev. B}\ }\textbf {\bibinfo {volume} {44}},\ \bibinfo {pages} {6199}
  (\bibinfo {year} {1991})}\BibitemShut {NoStop}%
\bibitem [{\citenamefont {Amman}\ \emph {et~al.}(1991)\citenamefont {Amman},
  \citenamefont {Wilkins}, \citenamefont {Ben-Jacob}, \citenamefont {Maker},\
  and\ \citenamefont {Jaklevic}}]{amman1991analytic}%
  \BibitemOpen
  \bibfield  {author} {\bibinfo {author} {\bibfnamefont {M}~\bibnamefont
  {Amman}}, \bibinfo {author} {\bibfnamefont {R}~\bibnamefont {Wilkins}},
  \bibinfo {author} {\bibfnamefont {E}~\bibnamefont {Ben-Jacob}}, \bibinfo
  {author} {\bibfnamefont {PD}~\bibnamefont {Maker}}, \ and\ \bibinfo {author}
  {\bibfnamefont {RC}~\bibnamefont {Jaklevic}},\ }\bibfield  {title} {\enquote
  {\bibinfo {title} {Analytic solution for the current-voltage characteristic
  of two mesoscopic tunnel junctions coupled in series},}\ }\href {\doibase
  10.1103/PhysRevB.43.1146} {\bibfield  {journal} {\bibinfo  {journal} {Phys.
  Rev. B}\ }\textbf {\bibinfo {volume} {43}},\ \bibinfo {pages} {1146}
  (\bibinfo {year} {1991})}\BibitemShut {NoStop}%
\bibitem [{\citenamefont {Delsing}\ \emph {et~al.}(1989)\citenamefont
  {Delsing}, \citenamefont {Likharev}, \citenamefont {Kuzmin},\ and\
  \citenamefont {Claeson}}]{delsing1989effect}%
  \BibitemOpen
  \bibfield  {author} {\bibinfo {author} {\bibfnamefont {P}~\bibnamefont
  {Delsing}}, \bibinfo {author} {\bibfnamefont {KK}~\bibnamefont {Likharev}},
  \bibinfo {author} {\bibfnamefont {Lo~S}\ \bibnamefont {Kuzmin}}, \ and\
  \bibinfo {author} {\bibfnamefont {T}~\bibnamefont {Claeson}},\ }\bibfield
  {title} {\enquote {\bibinfo {title} {Effect of high-frequency electrodynamic
  environment on the single-electron tunneling in ultrasmall junctions},}\
  }\href {\doibase 10.1103/PhysRevLett.63.1180} {\bibfield  {journal} {\bibinfo
   {journal} {Phys. Rev. Lett.}\ }\textbf {\bibinfo {volume} {63}},\ \bibinfo
  {pages} {1180} (\bibinfo {year} {1989})}\BibitemShut {NoStop}%
\bibitem [{\citenamefont {Devoret}\ \emph {et~al.}(1990)\citenamefont
  {Devoret}, \citenamefont {Esteve}, \citenamefont {Grabert}, \citenamefont
  {Ingold}, \citenamefont {Pothier},\ and\ \citenamefont
  {Urbina}}]{devoret1990effect}%
  \BibitemOpen
  \bibfield  {author} {\bibinfo {author} {\bibfnamefont {Michel~H}\
  \bibnamefont {Devoret}}, \bibinfo {author} {\bibfnamefont {Daniel}\
  \bibnamefont {Esteve}}, \bibinfo {author} {\bibfnamefont {Hermann}\
  \bibnamefont {Grabert}}, \bibinfo {author} {\bibfnamefont {G-L}\ \bibnamefont
  {Ingold}}, \bibinfo {author} {\bibfnamefont {Hugues}\ \bibnamefont
  {Pothier}}, \ and\ \bibinfo {author} {\bibfnamefont {Cristian}\ \bibnamefont
  {Urbina}},\ }\bibfield  {title} {\enquote {\bibinfo {title} {{Effect of the
  electromagnetic environment on the Coulomb blockade in ultrasmall tunnel
  junctions}},}\ }\href {\doibase 10.1103/PhysRevLett.64.1824} {\bibfield
  {journal} {\bibinfo  {journal} {Phys. Rev. Lett.}\ }\textbf {\bibinfo
  {volume} {64}},\ \bibinfo {pages} {1824} (\bibinfo {year}
  {1990})}\BibitemShut {NoStop}%
\bibitem [{\citenamefont {Brun}\ \emph {et~al.}(2012)\citenamefont {Brun},
  \citenamefont {M{\"u}ller}, \citenamefont {Hong}, \citenamefont {Patthey},
  \citenamefont {Flindt},\ and\ \citenamefont {Schneider}}]{brun2012dynamical}%
  \BibitemOpen
  \bibfield  {author} {\bibinfo {author} {\bibfnamefont {Christophe}\
  \bibnamefont {Brun}}, \bibinfo {author} {\bibfnamefont {Konrad~H}\
  \bibnamefont {M{\"u}ller}}, \bibinfo {author} {\bibfnamefont {I-Po}\
  \bibnamefont {Hong}}, \bibinfo {author} {\bibfnamefont {Fran{\c{c}}ois}\
  \bibnamefont {Patthey}}, \bibinfo {author} {\bibfnamefont {Christian}\
  \bibnamefont {Flindt}}, \ and\ \bibinfo {author} {\bibfnamefont
  {Wolf-Dieter}\ \bibnamefont {Schneider}},\ }\bibfield  {title} {\enquote
  {\bibinfo {title} {{Dynamical Coulomb blockade observed in nanosized
  electrical contacts}},}\ }\href {\doibase 10.1103/PhysRevLett.108.126802}
  {\bibfield  {journal} {\bibinfo  {journal} {Phys. Rev. Lett.}\ }\textbf
  {\bibinfo {volume} {108}},\ \bibinfo {pages} {126802} (\bibinfo {year}
  {2012})}\BibitemShut {NoStop}%
\bibitem [{\citenamefont {Senkpiel}\ \emph {et~al.}(2020)\citenamefont
  {Senkpiel}, \citenamefont {Kl{\"o}ckner}, \citenamefont {Etzkorn},
  \citenamefont {Dambach}, \citenamefont {Kubala}, \citenamefont {Belzig},
  \citenamefont {Yeyati}, \citenamefont {Cuevas}, \citenamefont {Pauly},
  \citenamefont {Ankerhold} \emph {et~al.}}]{senkpiel2020dynamical}%
  \BibitemOpen
  \bibfield  {author} {\bibinfo {author} {\bibfnamefont {Jacob}\ \bibnamefont
  {Senkpiel}}, \bibinfo {author} {\bibfnamefont {Jan~C}\ \bibnamefont
  {Kl{\"o}ckner}}, \bibinfo {author} {\bibfnamefont {Markus}\ \bibnamefont
  {Etzkorn}}, \bibinfo {author} {\bibfnamefont {Simon}\ \bibnamefont
  {Dambach}}, \bibinfo {author} {\bibfnamefont {Bj{\"o}rn}\ \bibnamefont
  {Kubala}}, \bibinfo {author} {\bibfnamefont {Wolfgang}\ \bibnamefont
  {Belzig}}, \bibinfo {author} {\bibfnamefont {Alfredo~Levy}\ \bibnamefont
  {Yeyati}}, \bibinfo {author} {\bibfnamefont {Juan~Carlos}\ \bibnamefont
  {Cuevas}}, \bibinfo {author} {\bibfnamefont {Fabian}\ \bibnamefont {Pauly}},
  \bibinfo {author} {\bibfnamefont {Joachim}\ \bibnamefont {Ankerhold}},  \emph
  {et~al.},\ }\bibfield  {title} {\enquote {\bibinfo {title} {{Dynamical
  Coulomb blockade as a local probe for quantum transport}},}\ }\href {\doibase
  10.1103/PhysRevLett.124.156803} {\bibfield  {journal} {\bibinfo  {journal}
  {Phys. Rev. Lett.}\ }\textbf {\bibinfo {volume} {124}},\ \bibinfo {pages}
  {156803} (\bibinfo {year} {2020})}\BibitemShut {NoStop}%
\bibitem [{\citenamefont {Anderson}(1959)}]{anderson1959theory}%
  \BibitemOpen
  \bibfield  {author} {\bibinfo {author} {\bibfnamefont {Philip~W}\
  \bibnamefont {Anderson}},\ }\bibfield  {title} {\enquote {\bibinfo {title}
  {Theory of dirty superconductors},}\ }\href {\doibase
  10.1016/0022-3697(59)90036-8} {\bibfield  {journal} {\bibinfo  {journal} {J.
  Phys. Chem. Solids}\ }\textbf {\bibinfo {volume} {11}},\ \bibinfo {pages}
  {26--30} (\bibinfo {year} {1959})}\BibitemShut {NoStop}%
\bibitem [{\citenamefont {Bose}\ \emph {et~al.}(2009)\citenamefont {Bose},
  \citenamefont {Galande}, \citenamefont {Chockalingam}, \citenamefont
  {Banerjee}, \citenamefont {Raychaudhuri},\ and\ \citenamefont
  {Ayyub}}]{bose2009competing}%
  \BibitemOpen
  \bibfield  {author} {\bibinfo {author} {\bibfnamefont {Sangita}\ \bibnamefont
  {Bose}}, \bibinfo {author} {\bibfnamefont {Charudatta}\ \bibnamefont
  {Galande}}, \bibinfo {author} {\bibfnamefont {SP}~\bibnamefont
  {Chockalingam}}, \bibinfo {author} {\bibfnamefont {Rajarshi}\ \bibnamefont
  {Banerjee}}, \bibinfo {author} {\bibfnamefont {Pratap}\ \bibnamefont
  {Raychaudhuri}}, \ and\ \bibinfo {author} {\bibfnamefont {Pushan}\
  \bibnamefont {Ayyub}},\ }\bibfield  {title} {\enquote {\bibinfo {title}
  {{Competing effects of surface phonon softening and quantum size effects on
  the superconducting properties of nanostructured Pb}},}\ }\href {\doibase
  10.1088/0953-8984/21/20/205702} {\bibfield  {journal} {\bibinfo  {journal}
  {J. Phys. Condens. Matter}\ }\textbf {\bibinfo {volume} {21}},\ \bibinfo
  {pages} {205702} (\bibinfo {year} {2009})}\BibitemShut {NoStop}%
\end{thebibliography}%

\end{document}